\documentclass{article}

\usepackage{PRIMEarxiv}

\usepackage[utf8]{inputenc} 
\usepackage[T1]{fontenc}    
\usepackage{hyperref}       
\usepackage{url}            
\usepackage{booktabs}       
\usepackage{amsfonts}       
\usepackage{nicefrac}       
\usepackage{microtype}      
\usepackage{lipsum}
\usepackage{fancyhdr}       
\usepackage{graphicx}       
\graphicspath{{media/}}     

\title{Billion-years old proteins show the importance of N-lobe orientation in Imatinib-kinase selectivity.
}

\author{
  Zahra Shamsi \\
  Department of Chemical and Biomolecular Engineering \\
  University of Illinois at Urbana-Champaign \\
  Urbana \\
  IL \\ 
   \And
  Diwakar Shukla \\
  Department of Chemical and Biomolecular Engineering \\
  University of Illinois at Urbana-Champaign \\
  Urbana \\
  IL \\ 
  \texttt{diwakar@illinois.edu} \\
}

\begin{document}
\maketitle

\begin{abstract}
The molecular origins of proteins' functions are a combinatorial search problem in the proteins' sequence space, which requires enormous resources to solve. However, evolution has already solved this optimization problem for us, leaving behind suboptimal solutions along the way. Comparing suboptimal proteins along the evolutionary pathway, or ancestors, with more optimal modern proteins can lead us to the exact molecular origins of a particular function.
In this paper, we study the long-standing question of the selectivity of Imatinib, an anti-cancer kinase inhibitor drug. We study two related kinases, Src and Abl, and four of their common ancestors, to which Imatinib has significantly different affinities. Our results show that the orientation of the N-lobe with respect to the C-lobe varies between the kinases along their evolutionary pathway and is consistent with Imatinib's inhibition constants as measured experimentally. The conformation of the DFG-motif (Asp-Phe-Gly) and the structure of the P-loop also seem to have different stable conformations along the evolutionary pathway, which is aligned with Imatinib's affinity.
\end{abstract}


\section{Introduction}
Protein kinases are a class of enzyme that transfer phosphate groups from ATP to other proteins, thereby signaling growth and cell proliferation. Mutations in kinases can lead to uncontrolled cell growth and eventually cancer, making kinases a prime target for drug design, typically of small molecule inhibitors \cite{Shamsi2018,Sawyers2002}. Imatinib is one of the clinically successful drugs for the treatment of multiple cancers like chronic myelogenous leukemia \cite{Capdeville2002}. It selectively inhibits Abl and not other structurally similar kinases like Src. Since the overactive Abl mutant only exists in cancer cells, Imatinib has a limited effect on healthy cells \cite{Seeliger2007}. Why does the drug inhibit Abl and not Src despite the high protein sequence identity between Src and Abl ($\sim$46\%) \cite{Seeliger2007}? After two decades of study on the basis of Imatinib selectivity between Abl and Src, the question still remains unanswered. During these years, extensive work has been done to elucidate this problem using different approaches from NMR and fast kinetics \cite{Agafonov2014}, sequence swapping \cite{Seeliger2007}, and ancestral gene reconstruction \cite{Wilson2015} experiments to free energy calculations and long time-scale molecular dynamics (MD) simulations \cite{Lin2013, Lin20130, Wang2018}. Each of these studies gave insight into different aspects of the problem, but a full answer is still missing.  

DFG-motif (Asp-Phe-Gly) is a highly conserved segment of the activation loop in kinase domains that is proposed to play a major role in the selection mechanism since Imatinib only binds a specific configuration of DFG. 
Two conformations of the DFG motif are the inactive ``DFG-out'' and the active ``DFG-in''. Imatinib only binds to DFG-out conformation of kinases. Multiple groups have argued that the kinetic basis of Imatinib selectivity for Abl, compared to Src kinase is because of a pre-existing equilibrium between DFG-out and DFG-in, or so-called a conformational selection mechanism \cite{Shan2008, Lin2013}. More recently despite the general belief on the critical role of conformational selection, Agafonov $et$ $al.$ claimed that the Imatinib selectivity is rooted in conformational changes after drug binding, not before \cite{Agafonov2014}. They support their hypothesis using NMR studies and showed the presence of induced-fit mechanism in Abl kinase. The next step was to find a sequence-function relationship, and clarify which set of residues is responsible for the accessibility of induced-fit mechanism in Abl, but not in Src. Sequence swapping had been performed to make Src similar to Abl but all studies had failed to illuminate the atomistic determinants of selectivity \cite{Seeliger2007, Lin2013}. On the other hand, phylogenetic studies have been used as powerful tools to study protein sequence-function relationships in different problems \cite{Harms2013, Gumulya2018, Paps2018}. Therefore, following the NMR study, Wilson $et$ $al.$ recreated the evolutionary pathway between Src and Abl by resurrecting the common ancestors between them \cite{Wilson2015}. Using x-ray structures of an ancestral kinase and binding kinetics data from NMR, they showed the evolution of their proposed induced-fit mechanism and its effect on the selectivity of Imatinib \cite{Wilson2015}.

However, proteins are not just sequences of amino acids; they can adopt thousands of different 3D conformations, among which only specific sets of conformations are functional. In the aforementioned studies with the plausible sequence-function relationship, still, the sequence-structure-function relationship is missing. Even though these studies have shown the presence of induced-fit mechanism, they do not give any details on what are these mechanisms in the protein's structure. 
In this study, we investigate the sequence-structure relationship by reconstructing the common ancestors of Src and Abl computationally.

\begin{figure}
\centering
\includegraphics[width=0.8\textwidth]{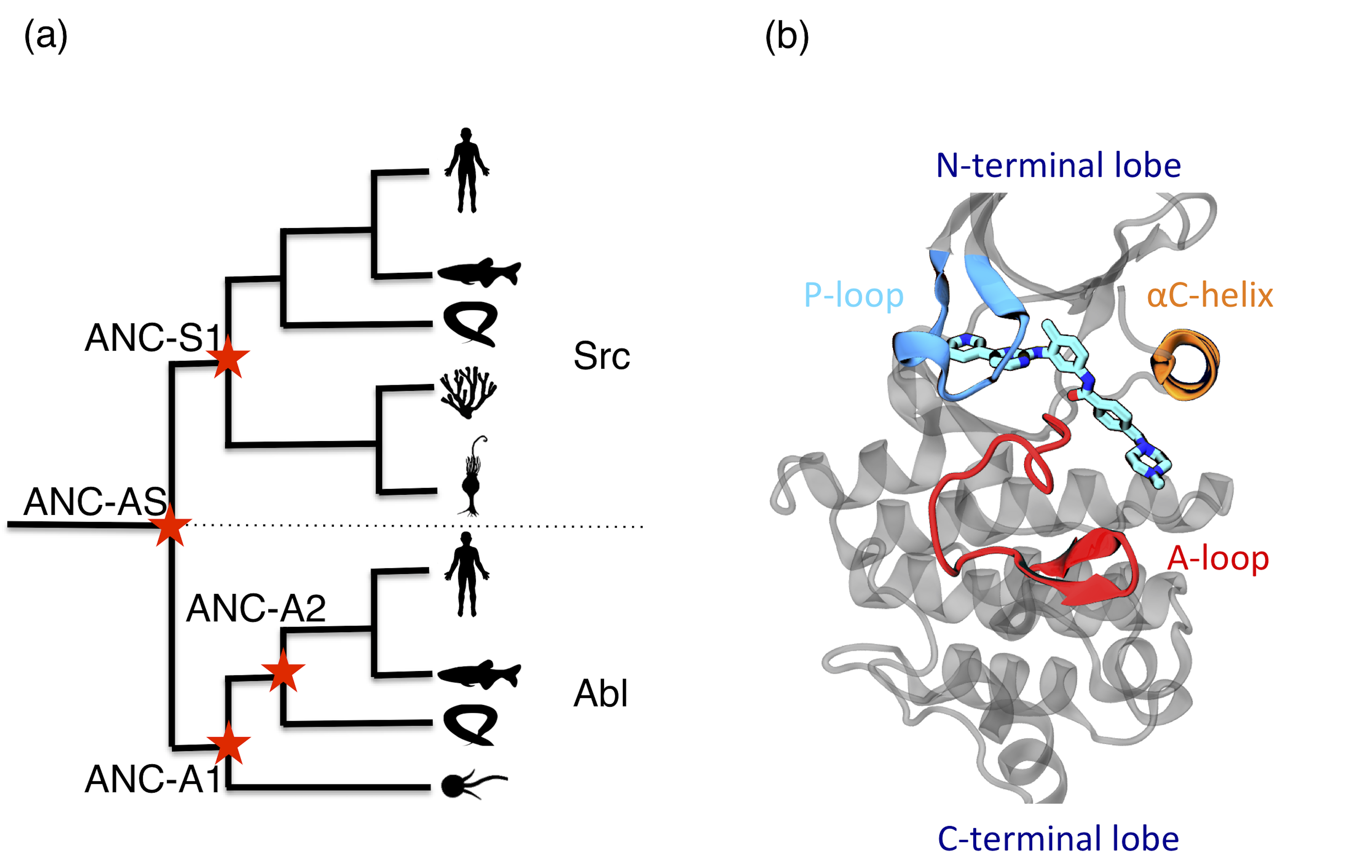}
\caption{\label{fig:tree}(a) Phylogenetic tree of Abl and Src families showing the reconstructed nodes.(b) Crystal structure of Abl kinase (PDB ID: 2HYY \cite{CowanJacob2006}) is shown with labels for important regions.}
\end{figure}

Study of evolutionary pathways is a natural way to identify the key amino acid changes that differentiate one family member from another \cite{Harms2010}. Evolution generates functional proteins in every stage despite the large sequence differences between them. It diversifies functions by altering their structure and the associated free-energy landscapes. The differences between Abl and Src also have evolved over a billion years from their common ancestor. Here, we borrow the sequences of the kinase ancestors from literature \cite{Wilson2015} and reconstruct them computationally (Figure \ref{fig:tree}(a) which is adapted from ref. \cite{Agafonov2015}). To elucidate the sequence-structure-function relationship, we simulated four ancestors, ANC-A1, ANC-A2, ANC-AS, and ANC-S1, plus ANC-AS with 15 suggested mutations which changed () the affinity significantly \cite{Wilson2015}. We refer to ANC-AS+15 as ANC-AS(+15) in this paper to prevent any confusion. We compare the results of these simulations with the experimental values of free energies and inhibition constants of the ancestors.
We also denote the evolution of conformations that play role in the conformational selection mechanism and induced-fit mechanism.

\section{Results}
\subsection{Evolution of conformational differences: N-lobe rotation}
Comparison of crystal structures shows a significant difference in the orientation of N-lobe with respect to the C-lobe between Src and Abl. This rotational angle is known to play a role in kinase activation, as the catalytic cleft between the lobes is relatively closed in inactive kinases \cite{Wilson2015}. The ancestral study by Wilson $et$ $al.$ suggested that altering only 15 residues in ANC-AS's N-lobe to the corresponding residues in Abl kinase drastically increased the Imatinib affinity to a level similar to Abl. This indicates the importance of N-lobe residues in the drug binding mechanism \cite{Wilson2015}. Therefore, it is likely that lobe rotation can be the dynamics effect of this 15 residue mutations in N-lobe. Here, we want to calculate the orientation of N-lobe along the evolutionary pathway and see if it is correlated with the Imatinib binding affinities.

\begin{figure}
\centering
\includegraphics[width=\textwidth]{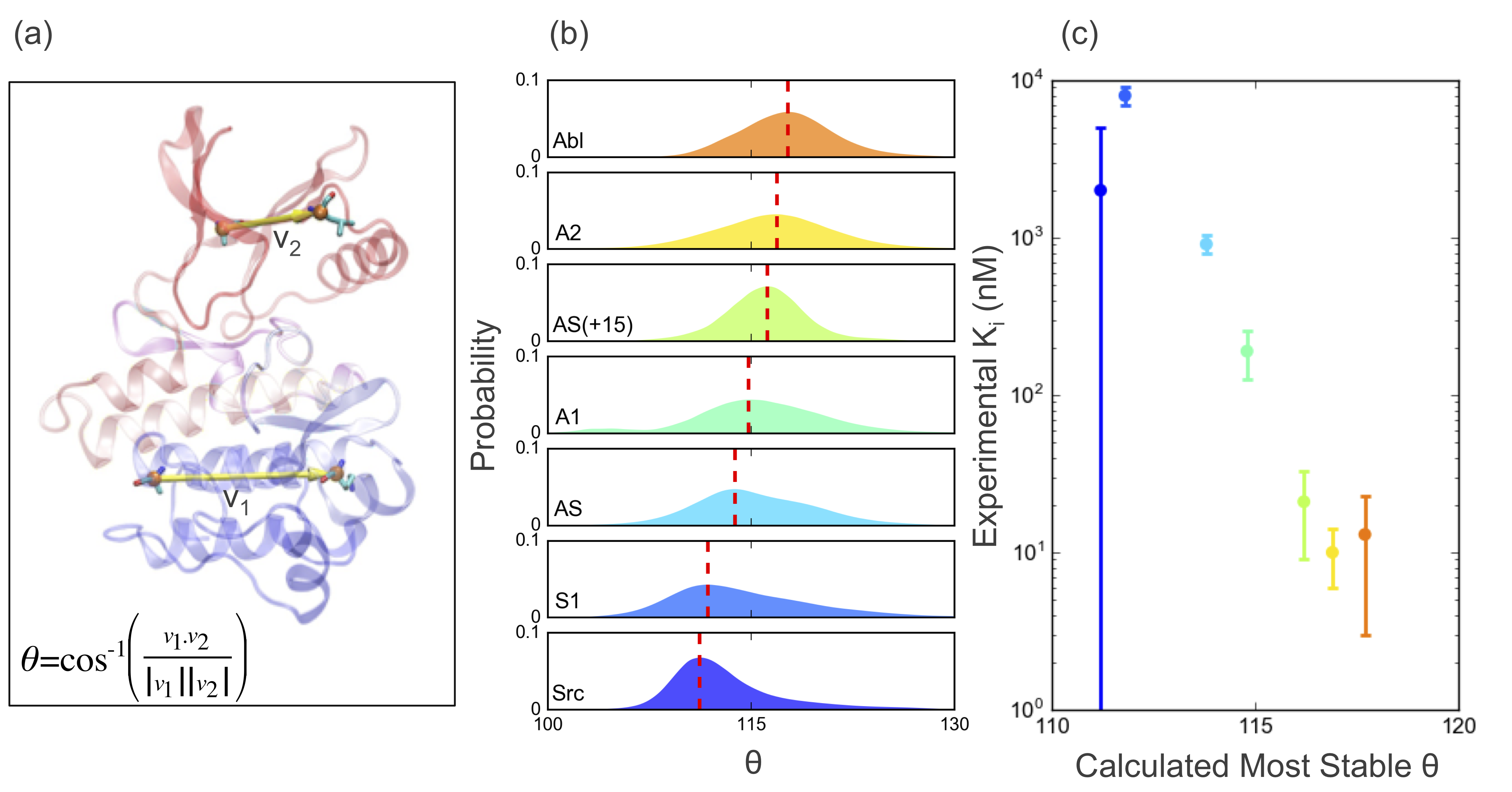}
\caption{\label{fig:theta} \textbf{N-lobe rotational angle of different kinases.} (a) The vectors defining the rotational angle are shown on the kinase structure. (b) The distribution of angles for different kinases reveals how position of N-lobe with respect to C-lobe evolved along the evolutionary pathway and the angle shifted from Src to Abl. (c) The correlation between the most likely value of $\theta$ calculated from the simulation and the experimental values of Imatinib inhibition constants ($K_i$) at 25\textdegree{}C as reported in literature \cite{Wilson2015}.}
\end{figure}

We studied the dynamics of four common ancestors of Src and Abl, an ancestor with the 15 mutations, and the two modern kinases. In these long timescale unbiased MD simulations no Imatinib molecule was present. The simulations were performed using adaptive sampling technique and Markov State Model (MSM) analysis \cite{EVsampling2017, Shukla2017, shamsi2018reinforcement}. In total, we performed 0.545 ms of aggregated unbiased MD simulation for the seven apo kinases. In order to quantify the lobe rotation, we define vectors in C-lobe (V1) and N-lobe (V2), and calculate the angle between them (named as $\theta$) as shown in Figure \ref{fig:theta}(a). Our results show a gradual shift in the density distributions of $\theta$ as we move from Abl to Src kinase on the phylogenetic tree (Figure \ref{fig:theta} (b)). The peak values of $\theta$ distributions are well correlated with the experimental values of Imatinib inhibition constants ($K_i$) at 25\textdegree{}C as reported in the literature \cite{Wilson2015}. This result suggests that the available area at the N-lobe and C-lobe is one of the major factors affecting the Imatinib binding affinity. 

\subsection{Evolution of induced-fit mechanism: secondary structure of P-loop.}
\begin{figure}
\centering
\includegraphics[width=0.9\textwidth]{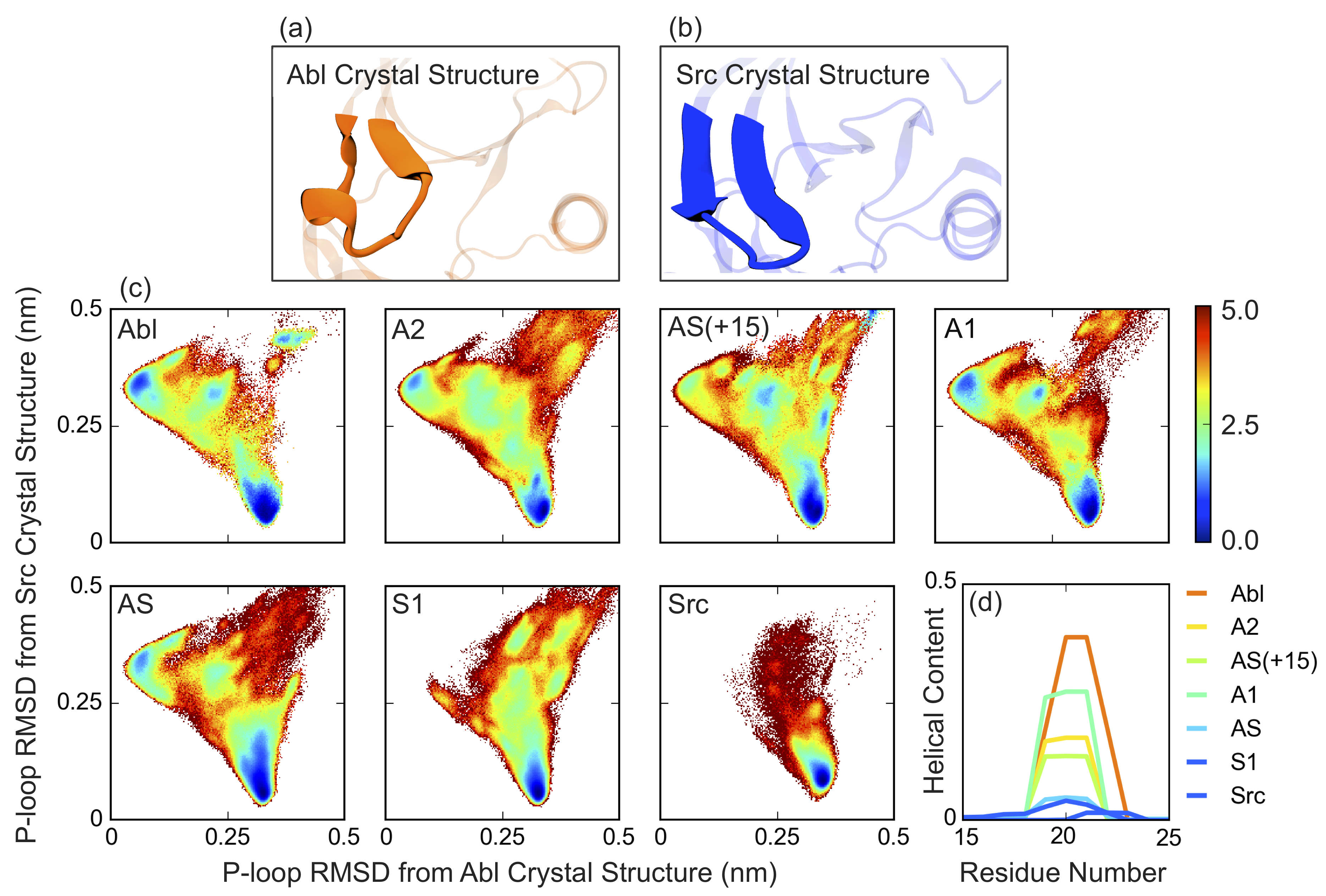}
\caption{\label{fig:ploop}\textbf{Evolutionary pathway of P-loop's conformation.} Secondary structure of P-loop in (a) Abl \cite{CowanJacob2006} and (b) Src's \cite{Seeliger2007} crystal structure is shown. Configuration of P-loop is the most distinct difference between Abl and Src's crystal structures. (c) Free energy landscapes of P-loop conformation show how ancestors lose their Abl-like P-loop structure as they get closer to Src. (d) Src and ancestors closer to Src and less likely to forms helical P-loop. Residue numbering is based on the sequence of ANC-AS(+15) as presented in the Figure S2. Colorbar shows free energies in kcal/mol.} 
\end{figure}

Changes in the secondary structure of P-loop are critical reaction coordinates for the Imatinib binding process. 
To observe the evolution of P-loop conformation, we calculated the root mean square deviation (RMSD) of the P-loop from the crystal structures of Abl and Src in the kinases (Figure \ref{fig:ploop}). Based on our observations, Src-like P-loop is stable in all seven kinases, whereas the ability to form Abl-like P-loop has been lost in ANC-S1 and Src. This suggests Imatinib binding from the P-loop side and the following induced-fit mechanism is not possible in Src and ANC-S1, but it could be feasible in the other five kinases. 
The exact residues responsible for Abl-like kinked P-loop should be a subset of the residue differences between ANC-AS and ANC-S1 (ANC-AS and ANC-S1 are 82\% identical as shown in Figure S1).

Looking at the simulations of Abl kinase and its crystal structure, we can observe that two inter residue interactions responsible for Abl P-loop kinked conformation are Y253-N322 and Q252-N322. Both of these interactions are conserved in ANC-A1, ANC-A2, and ANC-AS(+15), which makes P-loop more likely to form a helical structure as shown in Figure \ref{fig:ploop}(a). However, only one of the interactions is conserved in ANC-AS and ANC-S1, and none in Src. This partially explains the lower helical content of P-loop in ANC-AS, ANC-S1, and Src. The corresponding residue pairs are F-S and Q-S in ANC-AS/ANC-S1 and F-S and C-S in Src kinase.

\subsection{Evolution of conformational selection: DFG flip’s mechanism.}
\begin{figure}
\centering
\includegraphics[width=0.9\textwidth]{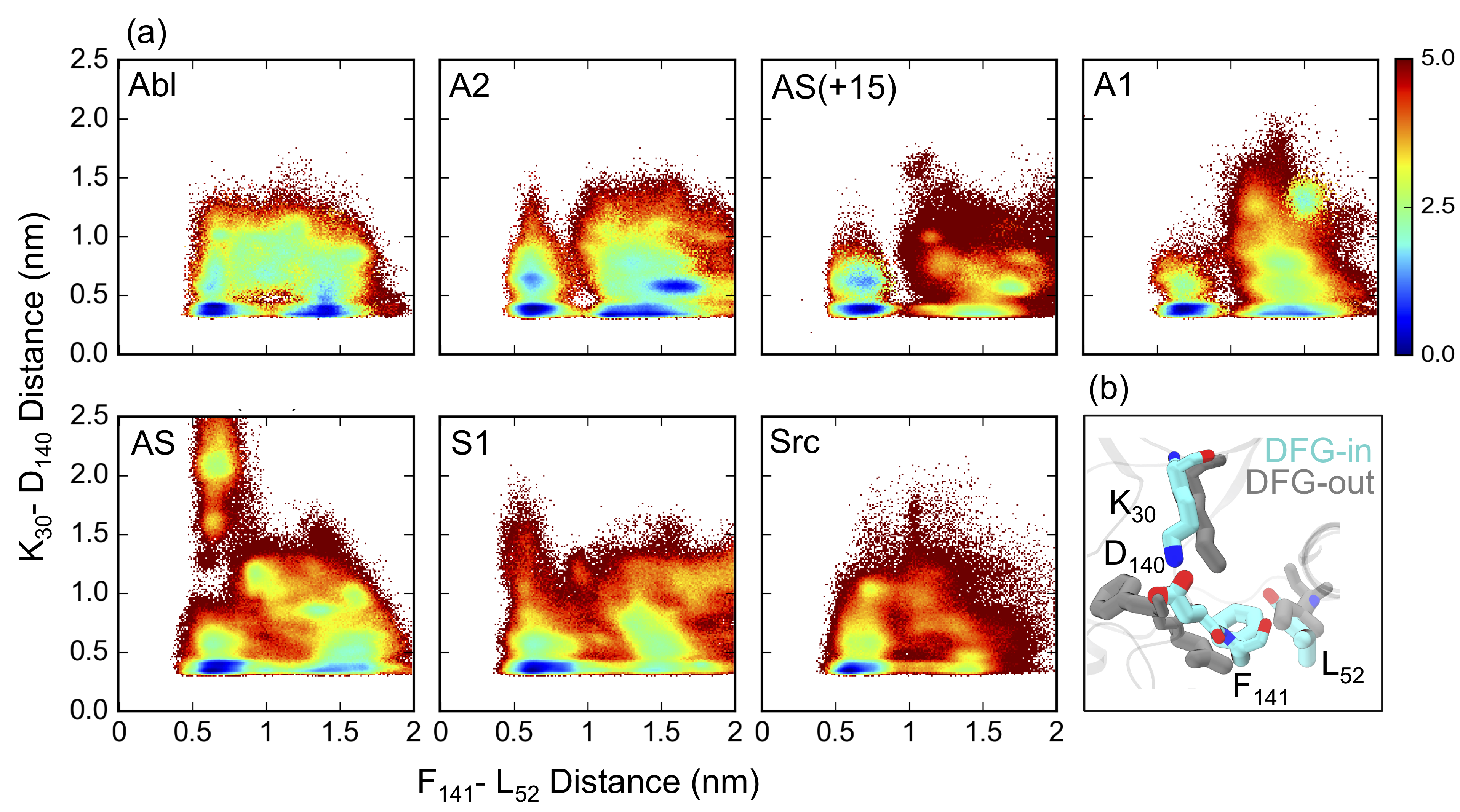}
\caption{\label{fig:dfg} \textbf{Stability of DFG-in and DFG-out conformations.} (a) Free energy landscapes of DFG flip for Abl, ancestors and Src are shown. When both the distances are greater than 1 nm, DFG motif is in DFG-out conformation, and when both of them are less than 1 nm it is in DFG-in conformation. Imatinib only binds DFG-out conformation. Colorbar shows free energies in kcal/mol. DFG-in and DFG-out conformations is shown in (b). Residue numbering is based on the sequence of ANC-AS(+15) as presented in the SI Figure S2}
\end{figure}

To investigate the dynamics of the DFG-motif, distances between two pairs of residues were measured as shown in Figure \ref{fig:dfg}. Our results indicate that all ancestors are able to adopt both DFG-in and DFG-out conformations. In Src kinase, DFG-out conformation is relatively less stable, whereas it is stable in Abl and all the ancestors. The difference in the stability of DFG-out conformation between Src and other kinases is $\sim$1 kcal/mol, which is not sufficient to justify the $\sim$3000 fold difference in their inhibition constants \cite{Seeliger2007}. Stable DFG-out conformation in the ancestors and Abl, indicates that the residues responsible for less stable DFG-out conformation in Src are among the set of residue differences between ANC-S1 and Src (ANC-S1 and Src are 75\% identical as shown in Figure S1).

\subsection{Evolution of conformational differences: secondary structure of A-loop.}
The activation-loop (A-loop) in a kinase is usually includes the highly conserved DFG motif and 17 of its following residues. This region adopts a closed or inactive conformation and an open or active conformation. In open conformations of A-loop, substrate proteins are able to bind the kinases for phosphate transfer, whereas in closed configurations, ATP is not exposed to the substrate proteins and phosphate transfer is not feasible. Imatinib binds the closed conformation of Abl and is incapable of binding to its open configuration \cite{la2002activity}.

\begin{figure}
\centering
\includegraphics[width=0.8\textwidth]{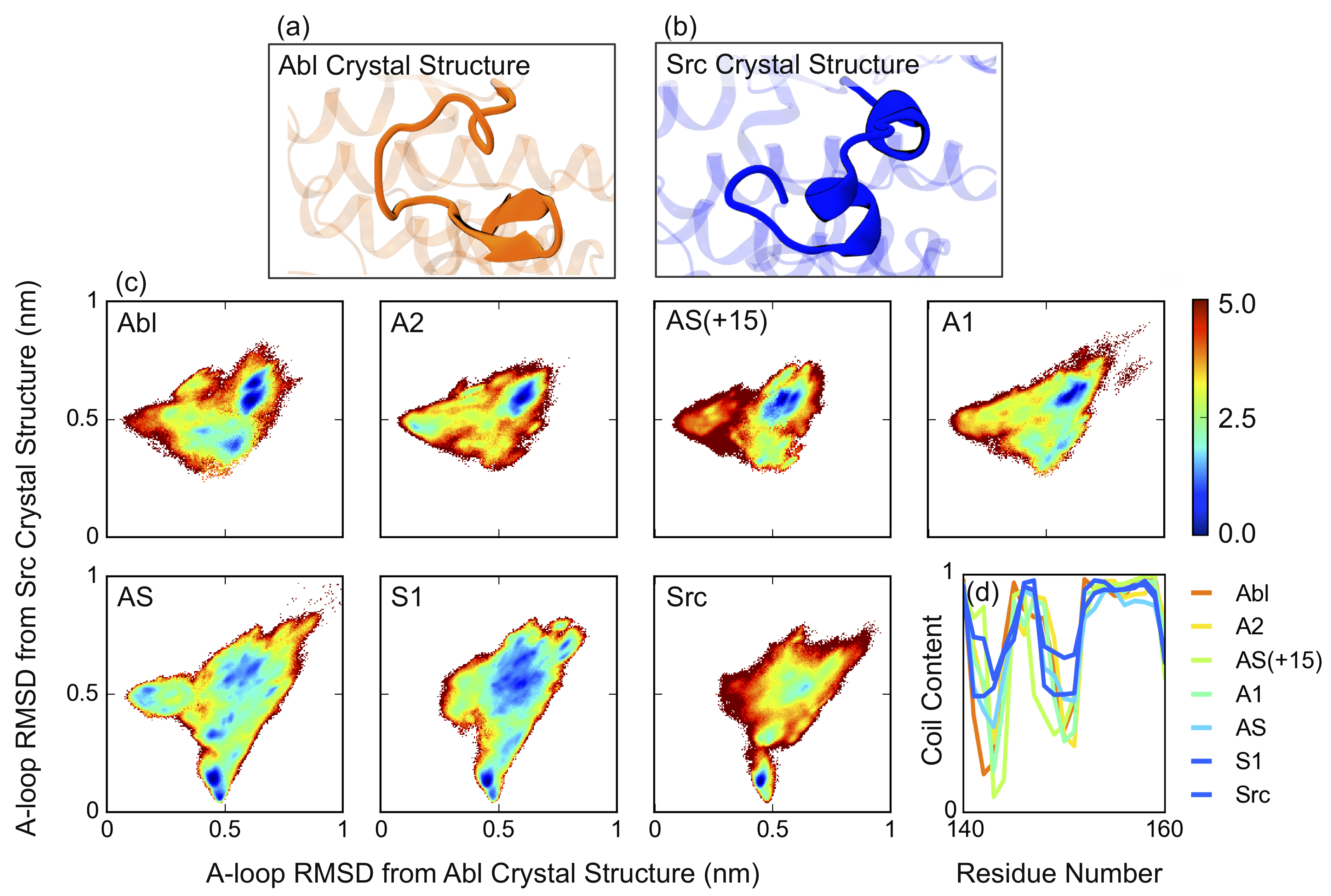}
\caption{\label{fig:aloop}\textbf{Evolutionary pathway of A-loop's conformation.} Secondary structure of A-loop in (a) Abl \cite{CowanJacob2006} and (b) Src \cite{Seeliger2007} crystal structure is shown. (c) Free energy landscapes of A-loop show how ancestors gain the ability to form Src-like A-loop structure as it gets closer to Src. Colorbar shows free energies in kcal/mol. (d) Src and ancestors closer to Src and less likely to $\beta$sheet A-loop. Residue numbering is based on the sequence of ANC-AS(+15).}
\end{figure}

Analyzing the simulations of the seven apo kinases showed that the Abl-like inactive configuration of A-loop is accessible in Abl's closer ancestors and disappears in ANC-S1. Therefore residues responsible for inactive Abl-like A-loop conformation should be among the sequence differences between ANC-S1 and ANC-AS (Figure \ref{fig:aloop}). On the other hand, Src-like inactive structure of A-loop appears in ANC-AS, which shows residues coding for Src-like helical conformation of A-loop are among the sequence differences between ANC-AS and ANC-AS(+15) (Figure \ref{fig:aloop}). We also see in Figure \ref{fig:aloop}(d) that A-loop in Src and ancestors closer to it tend to be less structured as compared to Abl.

\section{Discussion}
Protein-ligand selectivity remains a mysterious phenomenon in biology due to the lack of knowledge in biophysical mechanism of ligand binding. Here, we studied the long-lasting problem of Imatinib selectivity toward Abl kinase, using long MD simulations of Src, Abl, and their common ancestors. We compared the simulation results with experimental activity and binding free energy values to better understand the mechanism. We evaluated two of the best-known hypotheses that try to explain the selectivity mechanism by following their corresponding conformational changes along the evolutionary pathway connections Abl and Src kinases. The first hypothesis assumes the high selectivity of Imatinib towards Abl is due to its particular conformation of DFG motif. We compared the dynamics of DFG motif in Src, Alb and their ancestors and observed that all of them can adopt DFG-out conformation, which is crucial for Imatinib binding. However, DFG-out conformation seemed to be less stable in Src kinase as compared to the other six. This suggests that Imatinib can potentially bind to all of these kinases, but it is less likely to find the desirable conformation in Src kinase and bind to it. 
The second hypothesis that we studied was induced-fit mechanism. Even though we still do not know what exactly the induced-fit mechanism is, we believe P-loop conformation plays a role in the process. 
Therefore we studied the evolution of 3D structure of P-loop and observed that the helical P-loop adapts a $\beta$-sheet-rich conformation as we go from Src to Abl kinase on their phylogenetic tree. This change seemed to be very gradual along the evolutionary pathway and align with the Imatinib affinity towards the kinases measured by experiments \cite{Wilson2015}. 

Although Imatinib has achieved remarkable success in treating chronic myeloid leukemia, the emergence of resistance to this agent weakens the prospect of a cure for this leukemia. Imatinib resistance in most of the patients coincides with single point mutations in Abl kinase. In this study, we also rationalized the origin of some of these mutations.  

A set of mutations that reported to be among the most frequent mutation in Imatinib resistance patients are Y253F, Y253H, and Q252H in Abl kinase \cite{Cang2008}. We showed that these two residues are crucial to forming the helical conformation in the P-loop by making Y253-N322 and Q252-N322 interactions. 

In conclusion, our data suggest that future drug design efforts should focus more on understanding the binding pathways and their corresponding induced-fit steps. To this end, the full binding process of more drugs and drug targets need to be studied at atomic resolution with the hope that new insights will translate into improved next-generation compounds. As we learn more about Imatinib selectivity mechanism, these insights could be transferred to understand other selectivity challenges.

\bibliographystyle{unsrt}  
\bibliography{references}

\begin{thebibliography}{10}

\bibitem{Shamsi2018}
Zahra Shamsi and Diwakar Shukla.
\newblock How does evolution design functional free energy landscapes of
  proteins? a case study on the emergence of regulation in the cyclin dependent
  kinase family.
\newblock {\em Molecular Systems Design \& Engineering}, 5(1):392--400, 2020.

\bibitem{Sawyers2002}
Charles~L Sawyers.
\newblock Disabling abl{\textemdash}perspectives on abl kinase regulation and
  cancer therapeutics.
\newblock {\em Cancer Cell}, 1(1):13--15, February 2002.

\bibitem{Capdeville2002}
Renaud Capdeville, Elisabeth Buchdunger, Juerg Zimmermann, and Alex Matter.
\newblock Glivec ({STI}571, imatinib), a rationally developed, targeted
  anticancer drug.
\newblock {\em Nature Reviews Drug Discovery}, 1(7):493--502, July 2002.

\bibitem{Seeliger2007}
Markus~A. Seeliger, Bhushan Nagar, Filipp Frank, Xiaoxian Cao, M.~Nidanie
  Henderson, and John Kuriyan.
\newblock c-src binds to the cancer drug imatinib with an inactive abl/c-kit
  conformation and a distributed thermodynamic penalty.
\newblock {\em Structure}, 15(3):299--311, mar 2007.

\bibitem{Agafonov2014}
Roman~V Agafonov, Christopher Wilson, Renee Otten, Vanessa Buosi, and Dorothee
  Kern.
\newblock Energetic dissection of gleevec's selectivity toward human tyrosine
  kinases.
\newblock {\em Nature Structural {\&} Molecular Biology}, 21(10):848--853,
  September 2014.

\bibitem{Wilson2015}
C.~Wilson, R.~V. Agafonov, M.~Hoemberger, S.~Kutter, A.~Zorba, J.~Halpin,
  V.~Buosi, R.~Otten, D.~Waterman, D.~L. Theobald, and D.~Kern.
\newblock Using ancient protein kinases to unravel a modern cancer
  drug{\textquotesingle}s mechanism.
\newblock {\em Science}, 347(6224):882--886, February 2015.

\bibitem{Lin2013}
Y.-L. Lin, Y.~Meng, W.~Jiang, and B.~Roux.
\newblock Explaining why gleevec is a specific and potent inhibitor of abl
  kinase.
\newblock {\em Proceedings of the National Academy of Sciences},
  110(5):1664--1669, January 2013.

\bibitem{Lin20130}
Yen-Lin Lin and Beno{\^{\i}}t Roux.
\newblock Computational analysis of the binding specificity of gleevec to abl,
  c-kit, lck, and c-src tyrosine kinases.
\newblock {\em Journal of the American Chemical Society}, 135(39):14741--14753,
  September 2013.

\bibitem{Wang2018}
Lulu Wang, Guodong Zheng, Xianxian Liu, Duan Ni, Xinheng He, Jinying Cheng, and
  Shaoyong Lu.
\newblock Molecular dynamics simulations provide insights into the origin of
  gleevec's selectivity toward human tyrosine kinases.
\newblock {\em Journal of Biomolecular Structure and Dynamics},
  37(10):2733--2744, November 2018.

\bibitem{Shan2008}
Y.~Shan, M.~A. Seeliger, M.~P. Eastwood, F.~Frank, H.~Xu, M.~O Jensen, R.~O.
  Dror, J.~Kuriyan, and D.~E. Shaw.
\newblock A conserved protonation-dependent switch controls drug binding in the
  abl kinase.
\newblock {\em Proceedings of the National Academy of Sciences},
  106(1):139--144, December 2008.

\bibitem{Harms2013}
Michael~J. Harms and Joseph~W. Thornton.
\newblock Evolutionary biochemistry: revealing the historical and physical
  causes of protein properties.
\newblock {\em Nature Reviews Genetics}, 14(8):559--571, July 2013.

\bibitem{Gumulya2018}
Yosephin Gumulya, Jong-Min Baek, Shun-Jie Wun, Raine E.~S. Thomson, Kurt~L.
  Harris, Dominic J.~B. Hunter, James B. Y.~H. Behrendorff, Justyna Kulig, Shan
  Zheng, Xueming Wu, Bin Wu, Jeanette~E. Stok, James J.~De Voss, Gerhard
  Schenk, Ulrik Jurva, Shalini Andersson, Emre~M. Isin, Mikael Bod{\'{e}}n,
  Luke Guddat, and Elizabeth M.~J. Gillam.
\newblock Engineering highly functional thermostable proteins using ancestral
  sequence reconstruction.
\newblock {\em Nature Catalysis}, 1(11):878--888, October 2018.

\bibitem{Paps2018}
Jordi Paps and Peter W.~H. Holland.
\newblock Reconstruction of the ancestral metazoan genome reveals an increase
  in genomic novelty.
\newblock {\em Nature Communications}, 9(1), April 2018.

\bibitem{Harms2010}
Michael~J Harms and Joseph~W Thornton.
\newblock Analyzing protein structure and function using ancestral gene
  reconstruction.
\newblock {\em Current Opinion in Structural Biology}, 20(3):360--366, June
  2010.

\bibitem{Agafonov2015}
Roman~V. Agafonov, Christopher Wilson, and Dorothee Kern.
\newblock Evolution and intelligent design in drug development.
\newblock {\em Frontiers in Molecular Biosciences}, 2, May 2015.

\bibitem{CowanJacob2006}
Sandra~W. Cowan-Jacob, Gabriele Fendrich, Andreas Floersheimer, Pascal Furet,
  Janis Liebetanz, Gabriele Rummel, Paul Rheinberger, Mario Centeleghe, Doriano
  Fabbro, and Paul~W. Manley.
\newblock Structural biology contributions to the discovery of drugs to treat
  chronic myelogenous leukaemia.
\newblock {\em Acta Crystallographica Section D Biological Crystallography},
  63(1):80--93, dec 2006.

\bibitem{EVsampling2017}
Zahra Shamsi, Alexander Moffett, and Diwakar Shukla.
\newblock Enhanced unbiased sampling of protein dynamics using evolutionary
  coupling information.
\newblock {\em Sci. Rep.}, 5, sept 2017.
\newblock in-press.

\bibitem{Shukla2017}
Saurabh Shukla, Zahra Shamsi, Alexander~S. Moffett, Balaji Selvam, and Diwakar
  Shukla.
\newblock Application of hidden markov models in biomolecular simulations.
\newblock In {\em Hidden Markov Models}, pages 29--41. Springer New York, 2017.

\bibitem{shamsi2018reinforcement}
Zahra Shamsi, Kevin~J Cheng, and Diwakar Shukla.
\newblock Reinforcement learning based adaptive sampling: Reaping rewards by
  exploring protein conformational landscapes.
\newblock {\em The Journal of Physical Chemistry B}, 122(35):8386--8395, 2018.

\bibitem{la2002activity}
Paul La~Ros{\'e}e, Amie~S Corbin, Eric~P Stoffregen, Michael~W Deininger, and
  Brian~J Druker.
\newblock Activity of the bcr-abl kinase inhibitor pd180970 against clinically
  relevant bcr-abl isoforms that cause resistance to imatinib mesylate
  (gleevec, sti571).
\newblock {\em Cancer research}, 62(24):7149--7153, 2002.

\bibitem{Cang2008}
Shundong Cang and Delong Liu.
\newblock P-loop mutations and novel therapeutic approaches for imatinib
  failures in chronic myeloid leukemia.
\newblock {\em Journal of Hematology {\&} Oncology}, 1(1), October 2008.

\bibitem{Chojnacki2017}
Szymon Chojnacki, Andrew Cowley, Joon Lee, Anna Foix, and Rodrigo Lopez.
\newblock Programmatic access to bioinformatics tools from {EMBL}-{EBI} update:
  2017.
\newblock {\em Nucleic Acids Research}, 45(W1):W550--W553, apr 2017.

\bibitem{Robert2014}
Xavier Robert and Patrice Gouet.
\newblock Deciphering key features in protein structures with the new
  {ENDscript} server.
\newblock {\em Nucleic Acids Research}, 42(W1):W320--W324, apr 2014.

\end{thebibliography}

\pagebreak
\begin{center}
\textbf{\large Supplemental Materials}
\end{center}
\setcounter{equation}{0}
\setcounter{figure}{0}
\setcounter{table}{0}
\setcounter{page}{1}
\newcommand*\mycommand[1]{\texttt{\emph{#1}}}
\renewcommand{\thetable}{S\arabic{table}}
\renewcommand{\thefigure}{S\arabic{figure}}

\begin{table}[ht]
\centering
\begin{tabular}{ccc}
\hline
 & $K_i$ (nM) & $\Delta G$ (kcal/mol) \\
\hline
Abl     & 13 $\pm$ 10 & -4.5\\
ANC-A2  & 10 $\pm$ 4 & -5\\
ANC-AS(+15) & 21 $\pm$ 12 & -\\
ANC-A1  & 190 $\pm$ 65 & -3\\
ANC-AS & 910 $\pm$ 120 & -2.5\\
ANC-S1  & 8000 $\pm$ 1000 & -0.5\\
Src     & 2000 $\pm$ 3000 & -0.5\\
\hline
\end{tabular}
\caption{\label{tab:exper} Experimental values of Imatinib inhibition constants ($K_i$) at 25\textdegree{}C and $\Delta G$ associated with the induced-fit step from literature \cite{Wilson2015}.}
\end{table}

\clearpage
\begin{figure}
\centering
\includegraphics[width=\textwidth]{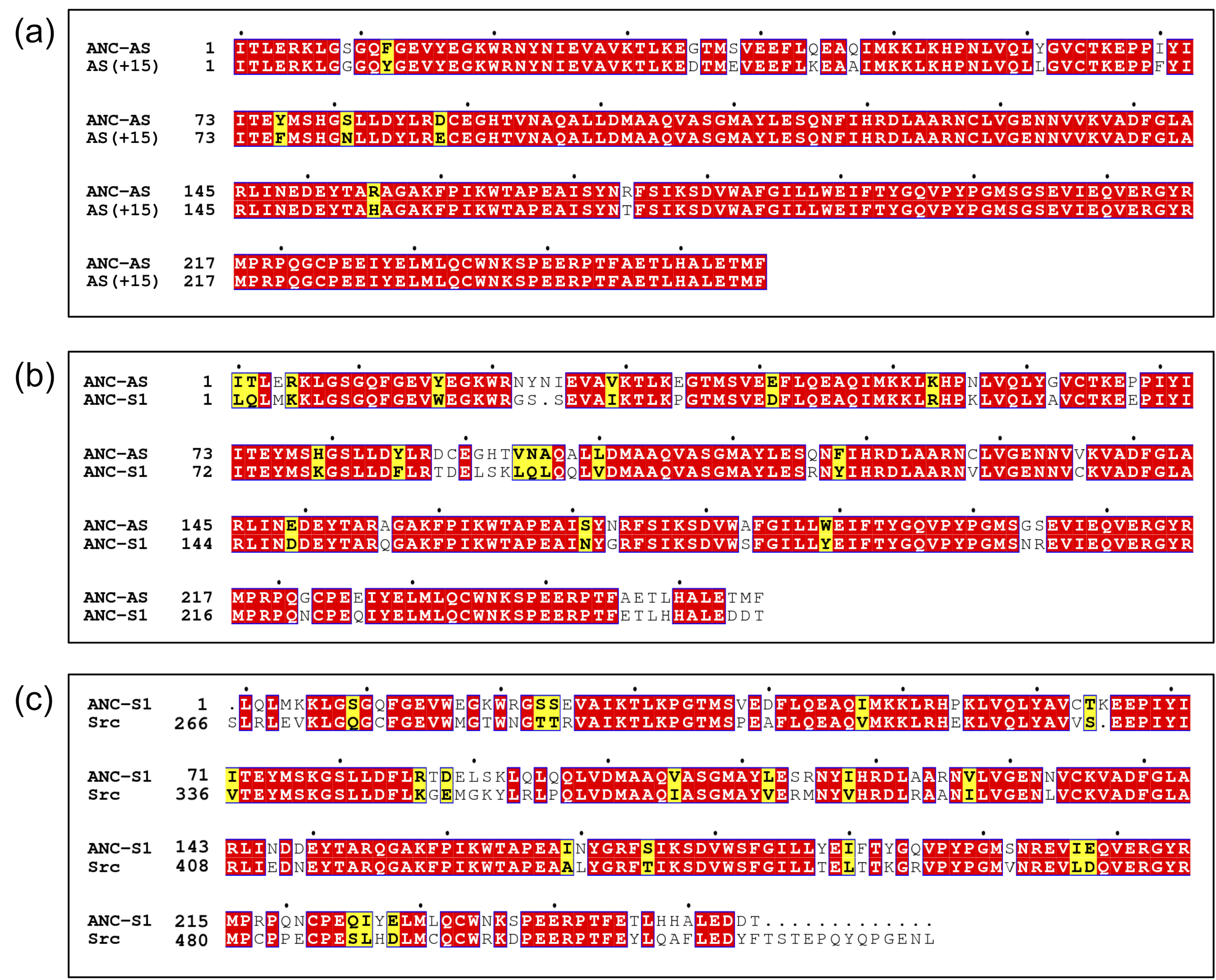}
\caption{\label{fig:s1}\textbf{Multiple sequence alignment of Src, Abl, and reconstructed ancestral sequences.} Identical and similar residues are boxed in red and yellow, respectively. Sequence alignments were created in T-Coffee \cite{Chojnacki2017} and plotted by ESPRIPT \cite{Robert2014}.
(a) ANC-AS and ANC-AS(+15) are 96\% identical. (b) ANC-AS and ANC-S1 are 82\% identical. (c) ANC-S1 and Src are 75\% identical.}
\end{figure}

\clearpage
\begin{figure}
\centering
\includegraphics[width=\textwidth]{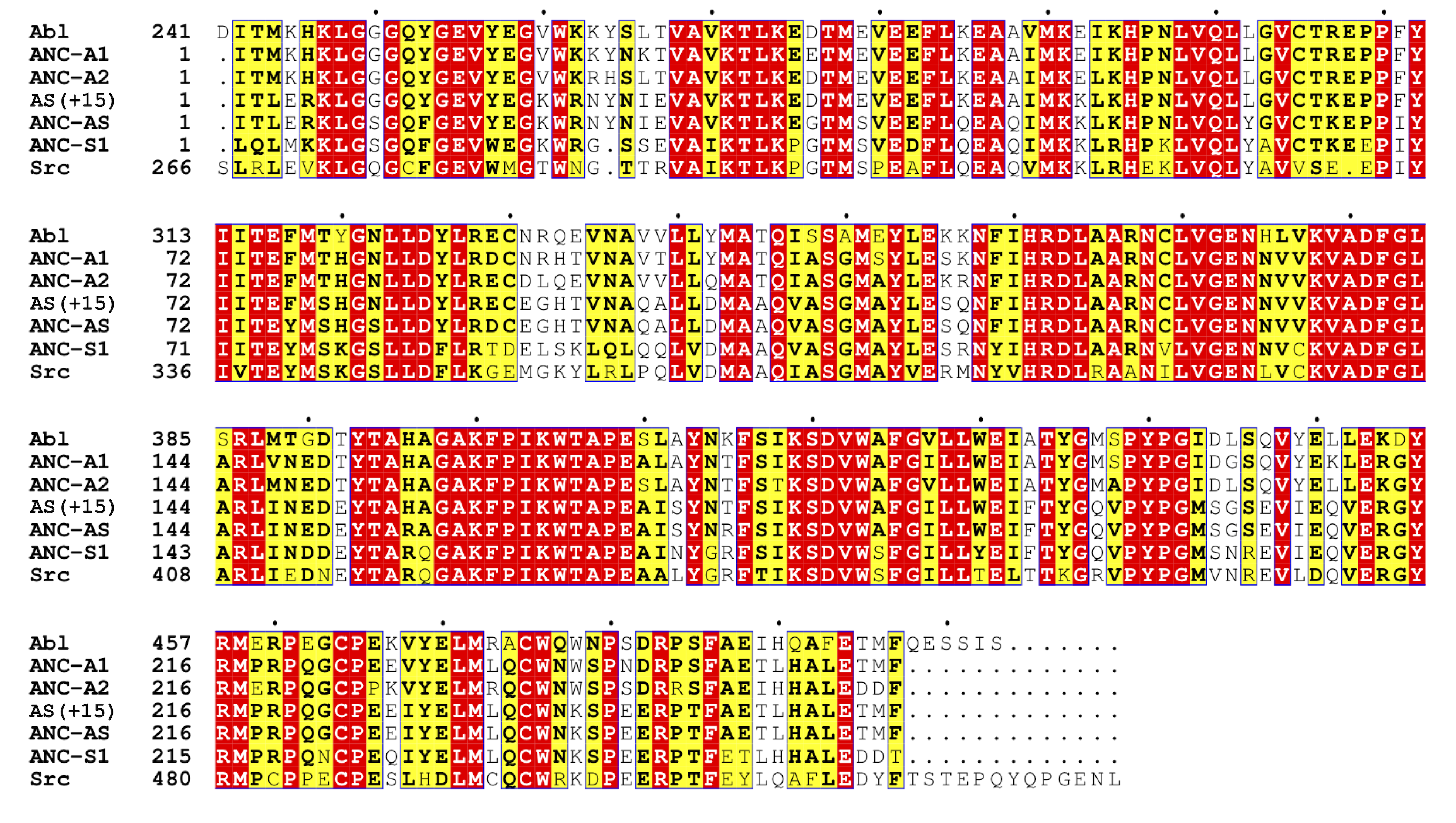}
\caption{\label{fig:s2}\textbf{Multiple sequence alignment of Src, Abl, and reconstructed ancestral sequences.}
Identical and similar residues are boxed in red and yellow, respectively. Sequence alignments were created in T-Coffee \cite{Chojnacki2017} and plotted by ESPRIPT \cite{Robert2014}.} 
\end{figure}

\clearpage
\begin{table}[ht]
\centering
\begin{tabular}{cccc}
\hline
 & N-lobe pocket interactions & A-loop pocket interactions & P-loop kinked Conf. \\
 & Y326L & F278Y & S346N  \\
\hline
Abl     & L & Y & N \\
ANC-A1  & L & Y & N\\
ANC-A2  & L & Y & N\\
AS(+15) & L & Y & N\\
ANC-AS & Y & F & S\\
ANC-S1  & Y & F & S\\
Src     & Y & F & S \\
\hline
\end{tabular}
\caption{\label{tab:mut} Conservation of some functionally important residues.}
\end{table}

\newpage

\begin{figure}
\centering
\includegraphics[width=\textwidth]{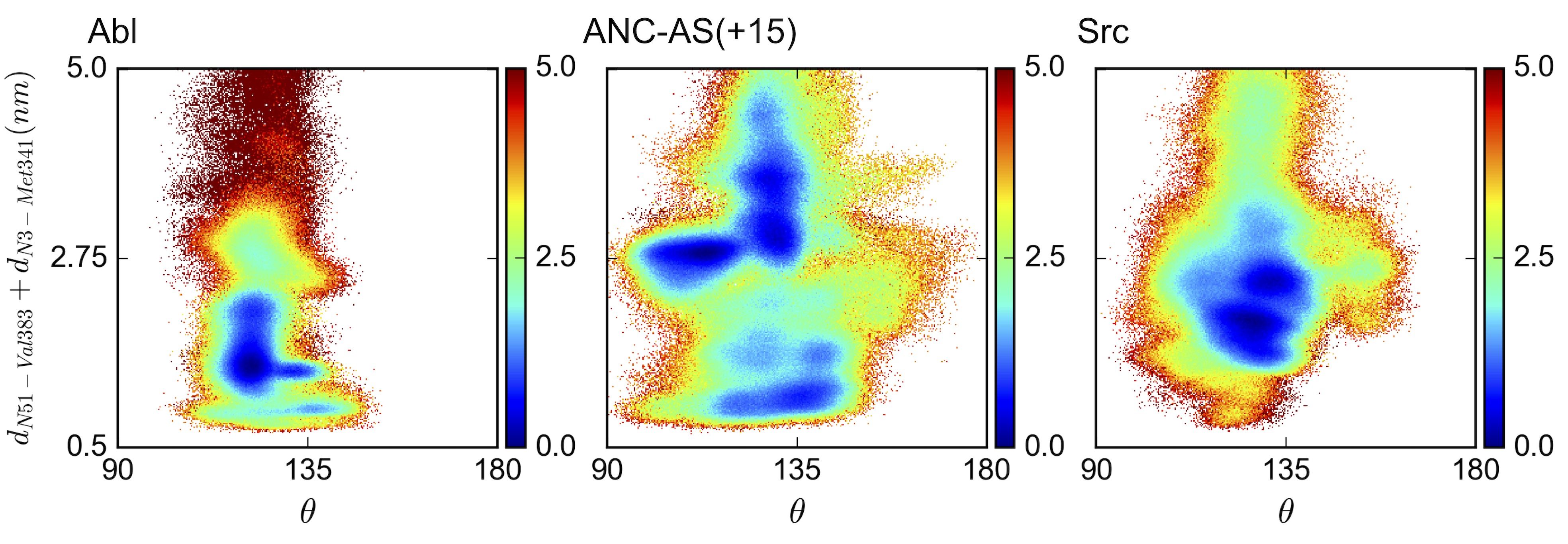}
\caption{\label{fig:s3}\textbf{The angle between two lobs as a function of Imatinib binding.} Imatinib binds the kinases when $\theta$ is equal or greater than 120\textdegree{}. Interaction with Imatinib increases the average value of the angles between the lobes.}
\end{figure}
\end{document}